\documentclass[prl,twocolumn,nofootinbib, preprintnumbers, superscriptaddress]{revtex4}

\usepackage{amsmath,amssymb,slashed}
\usepackage{graphicx}
\usepackage{epstopdf}
\usepackage{float}
\usepackage[colorlinks=true,
            linkcolor=blue,
            urlcolor=blue,
            citecolor=green,          
            bookmarks=true,
            bookmarksnumbered=true,
            breaklinks=true,
            pdfpagemode=Fullscreen,
            pdfstartview=FitBH]{hyperref}

\usepackage[normalem]{ulem}
\usepackage{color}

\definecolor{Orange}{cmyk}{0,0.61,0.87,0}
\definecolor{JungleGreen}{cmyk}{0.99,0,0.52,0}
\definecolor{OliveGreen}{cmyk}{0.64,0,0.95,0.40}
\definecolor{Brown}{cmyk}{0,0.81,1,0.60}
\definecolor{RoyalBlue}{cmyk}{0.71,0.53,0,0.12}
\definecolor{Gray}{cmyk}{0,0,0,0.40}
\definecolor{LightPink}{cmyk}{0.0,0.25,0,0}
\definecolor{LLightPink}{cmyk}{0.0,0.10,0,0}
\definecolor{LightBlue}{cmyk}{0.25,0,0,0}
\definecolor{LightGray}{cmyk}{0,0,0,0.2}

\usepackage{xcolor}
\definecolor{gesfpurple}{rgb}{0.47,0.19,0.42}

\definecolor{gesflanse}{rgb}{0.00,0.50,0.50}

\definecolor{gesfblue}{rgb}{0.08,0.42,0.76}

\definecolor{gesfred}{rgb}{1,0,0}

\definecolor{gesfwhite}{rgb}{1,1,1}

\definecolor{gesfblack}{rgb}{0,0,0}

\newcommand{\geqn}[1]{Eq.\,\hypersetup{linkcolor=blue}(\ref{#1})\hypersetup{linkcolor=blue}}
\newcommand{\gfig}[1]{{\hypersetup{linkcolor=violet}Fig.\,\ref{#1}\hypersetup{linkcolor=blue}}}

\graphicspath{{figures/}}

\begin{document}

\title{Unique Probe of Neutrino Electromagnetic Moments with Radiative Pair Emission}
\author{Shao-Feng Ge}
\email{gesf@sjtu.edu.cn}
\affiliation{Tsung-Dao Lee Institute \& School of Physics and Astronomy, Shanghai Jiao Tong University, China}
\affiliation{Key Laboratory for Particle Astrophysics and Cosmology (MOE) \& Shanghai Key Laboratory for Particle Physics and Cosmology, Shanghai Jiao Tong University, Shanghai 200240, China}
\author{Pedro Pasquini}
\email{ppasquini@sjtu.edu.cn}
\affiliation{Tsung-Dao Lee Institute \& School of Physics and Astronomy, Shanghai Jiao Tong University, China}
\affiliation{Key Laboratory for Particle Astrophysics and Cosmology (MOE) \& Shanghai Key Laboratory for Particle Physics and Cosmology, Shanghai Jiao Tong University, Shanghai 200240, China}

\begin{abstract}
The neutrino magnetic and electric moments are zero at tree level but
can arise in radiative corrections. Any deviation from the
Standard Model prediction would provide another indication
of neutrino-related new physics in addition to the neutrino
oscillation and masses. Especially, Dirac and Majorana neutrinos
have quite different structures in their electromagnetic moments.
Nevertheless, the recoil measurements and astrophysical
stellar cooling can only constrain combinations of neutrino
magnetic and electric moments with the limitation of not seeing
their detailed structures. We propose using the atomic radiative
emission of neutrino pair to serve as a unique probe of the neutrino
electromagnetic moments with the advantage of not just separating
the magnetic and electric moments but also identifying their
individual elements. Both searching strategy and projected
sensitivities are illustrated in this letter.
\end{abstract}

\maketitle 

{\it Introduction} --
In the Standard Model (SM) of particle physics, there is no 
tree-level coupling between neutrino ($\nu$) and photon ($A$)
\cite{PDG20}. However, the neutrino electromagnetic interactions 
are expected to arise from radiative corrections
\cite{Giunti:2014ixa},
\begin{eqnarray}
  H_M
=
\bar \nu
\left[
- f_M(q^2)i \sigma_{\mu\nu} q^\nu 
+ f_E(q^2)\sigma_{\mu \nu} q^\nu \gamma_5
\right]
  \nu A^\mu(q).
\,
\label{eq:nuMM}
\end{eqnarray}
The two terms account for the magnetic ($\mu_\nu \equiv 
f_M(0)$) and electric ($\epsilon_\nu \equiv f_E(0)$) 
dipole moments at vanishing momentum transfer, $q^2 = 0$, 
respectively.

With three neutrinos, both $\mu_\nu$ and $\epsilon_\nu$
are $3\times3$ hermitian matrices. For Majorana neutrinos,
their electromagnetic moments are 
antisymmetric under permutation, $(\mu_\nu)_{ij} = - 
(\mu_\nu)_{ji}$ and $(\epsilon_\nu)_{ij} = - 
(\epsilon_\nu)_{ji}$, implying that only the off-diagonal 
transition moments exist \cite{Schechter:1981hw,
Nieves:1981zt,Shrock:1982sc,Pal:1981rm,
Kayser:1982br,Kayser:1984ge}.
Observing a non-zero diagonal element $(\mu_\nu)_{ii}$
or $(\epsilon_\nu)_{ii}$ is then a direct evidence of
Dirac neutrinos. In addition, a nonzero diagonal $(\epsilon_\nu)_{ii}$
also indicates CP violation.

Generally speaking, the
explicit form of $\mu_\nu$ is model-depended and its
size is many orders smaller than the Bohr magneton
$\mu_B = e/2m_e$ 
\cite{Fujikawa:1980yx,Lee:1977tib,Petcov:1976ff,
Pal:1981rm,Shrock:1982sc,Bilenky:1987ty}. 
Interestingly, if the neutrino mass and magnetic moment arise 
from the same effective operator, the magnetic moment for
Majorana neutrino is typically 5 orders larger than
the Dirac one \cite{Bell:2005kz,Bell:2006wi}. Currently,
the experimental sensitivity is already around the threshold
for discovering the Majorana neutrino magnetic moments.

The neutrino electromagnetic moments can be tested in various
ways. Typically, the neutrino scattering cross section with
electron peaks in the low momentum transfer region
to provide a sizable signal in the electron recoil. Both solar
and reactor neutrinos can be used for such recoil measurement.
The best sensitivity comes from the reactor experiment
GEMMA, $\mu^{\rm eff}_{\alpha} < 2.9\times 10^{-11}\mu_B$ 
\cite{Beda:2013mta}, and the solar experiment Borexino,
$\mu^{\rm eff}_{\alpha} <2.8\times10^{-11}\mu_B$
\cite{Borexino:2017fbd}, both at 90\% C.L. However, the recoil 
measurement probes not just the magnetic moment $\mu_\nu$ but also the
electric one $\epsilon_\nu$ as a combination
\cite{Grimus:1997aa,Beacom:1999wx},
\begin{eqnarray}
   (\mu^{\rm eff}_{\alpha})^2 
\equiv
  \sum_j
\left| \sum_k U_{\alpha k}^*
  \left[ 
    (\mu_{\nu})_{jk} - i(\epsilon_{\nu})_{jk}
  \right]
\right|^2,
\label{eq:nu_magnetic_moment_effective}
\end{eqnarray}
where $U_{\alpha k}$ is the neutrino mixing matrix element
\cite{PDG20}. 
The sensitivity of scattering experiments only applies
to the combination $\mu^{\rm eff}_{\alpha}$ but not
the individual electromagnetic moments due to possible
cancellation among them. In other words, 
there is no unique probe of the neutrino magnetic or electric moment.

Similarly, the stellar cooling due to plasmon decay
($\gamma^*\rightarrow \bar \nu \nu$) is also sensitive to
a combination \cite{Giunti:2014ixa},
\begin{eqnarray}
  (\mu^\odot_{\nu})^2
\equiv
  \sum_{ij}
  |(\mu_{\nu})_{ij}|^2 + |(\epsilon_{\nu})_{ij}|^2,
  \label{eq:magnetic_moment_stellar}
\end{eqnarray}
rather than an individual magnetic moment.
The current bounds are
$\mu_\nu^\odot<2.2\times10^{-12}\mu_B$ from red giants
\cite{Diaz:2019kim}, $\mu_\nu^\odot < 7\times 10^{-12}\mu_B$
from white dwarfs pulsation \cite{Corsico:2014mpa},
and $\mu_\nu^\odot < 2.9\times 10^{-12}\mu_B$ from
white dwarfs cooling \cite{MillerBertolami:2014oki,Hansen:2015lqa}
at 90\% C.L.
Even though astrophysical bounds are stronger than the
scattering counterparts, the stellar modelling contains various
systematic uncertainties
\cite{Stancliffe:2016sa}.

It is of interest to mention that
the dark matter (DM) direct detection experiments can also 
constrain $\mu^{\rm eff}$ via the solar neutrino scattering
with electron. In fact, the effective neutrino magnetic 
moment in \geqn{eq:nu_magnetic_moment_effective}  
explains the recent Xenon1T data excess if 
$\mu_{\alpha}^{\rm eff} \approx 2\times10^{-11}\mu_B$ 
\cite{XENON:2020rca} which is also consistent 
with the PandaX-II data \cite{PandaX-II:2020udv,PandaX-II:2021nsg}.
The excess has prompted many studies on the neutrino magnetic
properties including active-to-active  \cite{Miranda:2020kwy,Chala:2020pbn,Khan:2020vaf} 
and active-to-sterile magnetic 
moments \cite{Shoemaker:2020kji,Brdar:2020quo}.
Future DM direct detection experiments can further
improve the sensitivity
\cite{AristizabalSierra:2020zod,Ye:2021zso,Schwemberger:2022fjl,Li:2022bqr}.

Although having multiple experimental ways,
the current probe of neutrino electromagnetic moments has intrinsic 
limitations. In addition to the fact that the aforementioned measurements
cannot distinguish the magnetic moment from the electric counterpart,
the presence of the mixing matrix leads to blind spots in the allowed 
parameter space \cite{AristizabalSierra:2021fuc}.
Moreover, existing measurements can not truly probe the magnetic
moment at zero momentum transfer but instead have an $\mathcal 
O({\rm keV})$ threshold. It is desirable to find new ways 
of exploring the neutrino electromagnetic properties.

In this letter, we present a novel way to probe
the neutrino magnetic and electric moments by using the
proposed {\it radiative emission of neutrino pair} (RENP)
\cite{Yoshimura:2006nd,Fukumi:2012rn,Tashiro:2019ghs}.
Although the RENP transition is yet to be observed, the coherent superradiance has been demonstrated with
two-photon emission from hydrogen molecules \cite{Hiraki:2018jwu}. Further discussions on experimental realization and background suppression can be found in \cite{Yoshimura:2015fna,Tanaka:2016wir,Tanaka:2019blr}.

The RENP process with $\mathcal O({\rm eV})$ momentum transfer 
is a perfect place for probing light mediator interactions
\cite{Ge:2021lur}. With massless photon being the mediator,
the neutrino electromagnetic interactions fall exactly into this category.
It allows the possibility of scanning the detailed structure of neutrino magnetic and electric moments in the mass eigenstate basis as a unique probe.

{\it Electromagnetic Emission of Neutrino Pair} --
The radiative emission of a neutrino pair 
is an atomic transition from an excited 
state $|e \rangle$ to the ground state $|g\rangle$.
With the direct transition $|e\rangle\rightarrow 
|g\rangle + \gamma$ being forbidden, the emission 
arises at the second order in perturbation 
theory. The atom first goes from an excited state $|e\rangle$ 
to a virtual state $|v\rangle$ and then falls to the ground
state $|g\rangle$,
\begin{eqnarray}
  |e\rangle
\rightarrow
  |v\rangle + \bar \nu \nu
\rightarrow 
  |g\rangle + \gamma + \bar \nu \nu.
\end{eqnarray}
This spontaneous process is very slow, but can be 
greatly enhanced by superradiance using a trigger 
laser beam \cite{Yoshimura:2006nd,Yoshimura:2012tm}.

The total Hamiltonian describing the reaction contains three parts,
\begin{eqnarray}
  H 
=
  H_0 + D_\gamma + H_W.
\end{eqnarray}
The zeroth-order Hamiltonian, $H_0$, accounts for the 
electron state, 
$H_0 | a \rangle = E_a | a \rangle$ where
$a = v$, $e$, or $g$. With energies $E_v > E_e > E_g$, the two-step 
process $|e\rangle\rightarrow|v\rangle\rightarrow |g\rangle$ 
renders $|e\rangle$ meta-stable. By proper selection of
$| e \rangle$ and $| g \rangle$,
the whole transition is of M1$\times$E1 type with
one electric (E1) and one magnetic (M1) dipole transitions.

The photon is emitted from the second step, 
$|v \rangle \rightarrow | g \rangle + \gamma$, by the
E1-type electric dipole term $D_\gamma$ \cite{Song:2015xaa}.
The corresponding amplitude is,
\begin{eqnarray}
   \langle g|D_\gamma| v \rangle 
\equiv
  \mathcal M_D  e^{-i\omega t + {\bf k}\cdot {\bf x}},
\quad
  \mathcal M_D
\equiv
- {\bf d}_{gv}\cdot {\bf E}_0,
\label{eq:relations_atomic}
\end{eqnarray}
where $\omega$ and ${\bf k}$ are the photon energy
and momentum, respectively. The matrix element $\mathcal M_D$
is a product of the dipole operator ${\bf d}_{gv}$
for the atomic transition $|v\rangle\rightarrow|g\rangle$
and the photon electric field ${\bf E}_0$.

On the other hand, the neutrino pair emission
$|e\rangle\rightarrow|v\rangle + \bar \nu_j\nu_i$ during
the first step is of M1 type dictated by the weak Hamiltonian
$H_W$. In the SM, the leading contribution comes from the
electroweak (EW) charged and neutral currents,
\begin{subequations}
\begin{align}
  \langle v| H_W | e \rangle 
& =
  \mathcal M_W e^{-i(p_{\nu_i} + p_{\nu_j})\cdot x},
\\  
\mathcal M_W
& =
  - a_{ij}\sqrt{2} G_F
  \langle v|\bar e \gamma_\mu \gamma_5 e| e \rangle 
  \left(\bar u_{iL}\gamma_\mu v_{jL}\right),
\end{align}
\label{eq:MW}
\end{subequations}
where the prefactor $a_{ij} \equiv U_{ei}U_{ej}^* - \delta_{ij}/2$
is a function of the neutrino mixing matrix elements $U_{ei}$.
Although both vector and axial-vector currents are present,
only the axial part of the electron current contributes since
the transition is of the M1 type \cite{Song:2015xaa}.

\begin{figure}[t]
\centering
\includegraphics[width=0.22\textwidth]{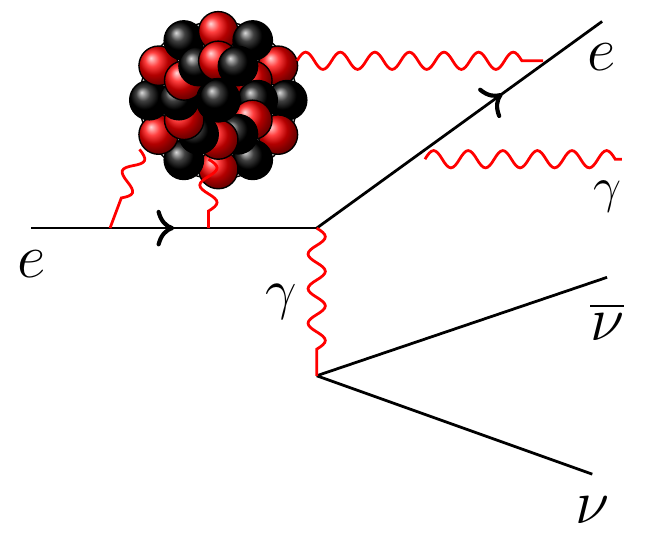}
\includegraphics[width=0.22\textwidth]{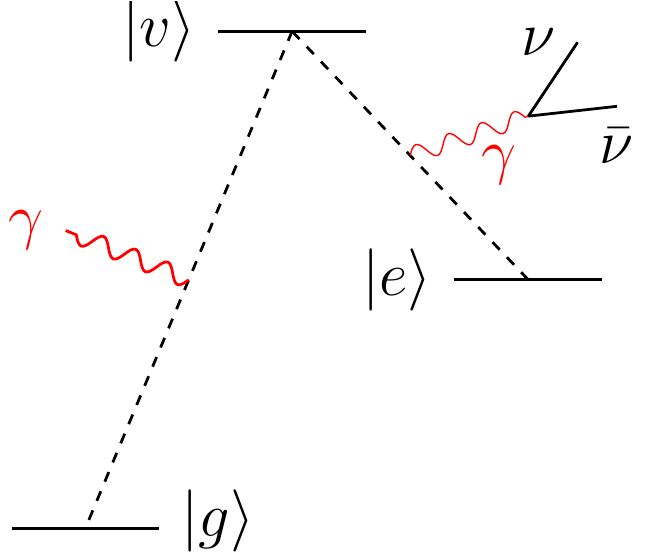}
\caption{
{\bf Left:} 
  The Feynman diagram for the RENP emission from a non-zero neutrino magnetic moment.
{\bf Right :} 
  The atomic transition diagram for the RENP
  process. In a typical experimental configuration,
  the emission of a neutrino pair is emitted first from
  $|e\rangle \rightarrow |v\rangle + \bar \nu \nu$ while the photon
  emission occurs for $|v\rangle \rightarrow |g\rangle + \gamma$.
}
\label{fig:EM_RENP_process}
\end{figure}

Non-zero neutrino magnetic and electric moments
in \geqn{eq:nuMM} can also contribute to the M1 type
transition as depicted in \gfig{fig:EM_RENP_process}. 
For Dirac neutrinos, the amplitude for the magnetic one is
$\langle v| H_M | e \rangle =
  \mathcal M_M e^{-i(p_{\nu_i} + p_{\nu_j})\cdot x}$
with
\begin{eqnarray}
  \mathcal M_M
=
  \mu_B (\mu_\nu)_{ij}
  \frac{ q_\nu q^\beta}{q^2}
\langle v| \overline e\sigma^{\mu \nu} e | e \rangle
\bar u_{i} \sigma_{\beta \mu} v_j,
\end{eqnarray}
while the electric one 
$\langle v| H_E | e \rangle =
  \mathcal M_E e^{-i(p_{\nu_i} + p_{\nu_j})\cdot x}$
has
\begin{eqnarray}
  \mathcal M_E
=
  \mu_B (\epsilon_\nu)_{ij}
  \frac{ q_\nu q^\beta}{q^2}
\langle v| \overline e\sigma^{\mu \nu} e | e \rangle
\bar u_{i} \sigma_{\beta \mu} \gamma_5 v_j.
\label{eq:EM_matrix}
\end{eqnarray}
The momentum transfer is defined as $q \equiv p_{\nu_i} + p_{\nu_j} 
= (E_{eg} - \omega, - {\bf k})$ with $E_{eg} \equiv E_e - E_g$. 
For Majorana neutrinos, there is an extra contribution,
\begin{align}
  \mathcal M_M^{(M)}
& =
  \mu_B \frac{q_\nu q^\beta}{q^2}
  \langle v|
    \overline e\sigma^{\mu \nu} 
    e
    | e \rangle
\nonumber
\\
& \times
    \frac{
        (\mu_\nu)_{ij}
        \bar u_{i}
        \sigma_{\beta \mu} 
        u_j
    -
        (\mu_\nu)_{ji}
        \bar u_{j}
        \sigma_{\beta \mu} 
        v_i
      }{2}.
\end{align}
and similarly for the electric moment case with the vertex replacement
$(\mu_\nu)_{ij}\sigma_{\beta \mu}\rightarrow (\epsilon_\nu)_{ij}
\sigma_{\beta \mu}\gamma_5$.
The minus sign comes from the anti-commutation property 
of fermion fields and the factor of $1/2$ from the Lagrangian
interaction normalization of Majorana neutrinos.
Using $(\mu_\nu)_{ji} = - (\mu_\nu)_{ij}$
and $\bar u_{j} \sigma_{\beta \mu} v_i = - \bar u_{i} \sigma_{\beta 
\mu} v_j$, the Majorana case is the same as its Dirac counterpart,
$\mathcal M_M^{(M)} = \mathcal M_M$
for $i\neq j$. It is also true for the electric moment case,
$\mathcal M_E^{(M)} = \mathcal M_E$.

For non-relativistic atomic states, only the spatial
components of the atomic currents $\langle v|\overline e 
\gamma^\mu \gamma_5 e | e \rangle$ in \geqn{eq:MW} and
$\langle v|\overline e\sigma^{\mu\nu} e| e \rangle$
in \geqn{eq:EM_matrix} contribute significantly.
They are proportional to the atomic spin operator ${\bf S}$ 
\cite{Weinberg:1995mt},
$\langle v| \overline e \boldsymbol{\gamma} \gamma_5 e | e \rangle
=
  2 {\bf S}_{ve}$ and
$\langle v| \overline e\sigma^{ij}  e | e \rangle
=
- 2 \epsilon_{ijk}{\bf S}_{ve}^k$.
The summation over the electron spins $m_e$ and $m_v$
follows the identity \cite{Dinh:2012qb},
\begin{eqnarray}
  \label{eq:}
  \frac{1}{2(2J_e + 1)}
   \sum_{m_e m_v}
   {\bf S}^i_{ve}
   {\bf S}^j_{ve}
=
  (2 J_v + 1) \frac{C_{ve}} 3 \delta_{ij}.
\end{eqnarray}
The other factors $J_a$ are the total spin of the excited ($a = e$)
and virtual ($a = v$) states. For Yb and Xe, 
$(2 J_v + 1)C_{ve} = 2$ \cite{Song:2015xaa}.

In addition to the SM contribution $|\mathcal M_W|^2$
\cite{Fukumi:2012rn,Song:2015xaa,Ge:2021lur},
the neutrino magnetic/electric moment first contributes
an spin averaged term,
\begin{eqnarray}
  \overline{|\mathcal M_{\overset M E}|^2}
=
  \frac{8 C_{ev}} 3 (2J_v + 1)
  \frac{\mu_B^2\omega^2}{q^4}
\times
  \left\{ |(\mu_\nu)_{ij}|^2, (\epsilon_\nu)_{ij}|^2 \right\}
\nonumber
\\
 \times
\left[
  q^2 (m_i \pm m_j)^2 
- (\Delta m_{ji}^2)^2
+ 2 q^2 |{\bf p}_{\nu_i}|^2 \sin^2 \theta
\right],
\quad 
\label{eq:Msqrd_final}
\end{eqnarray}
where $\theta$ is the angle between the photon and the
neutrino momentum. Between the magnetic and electric
moments, the mass eigenvalue $m_j$ flips a sign which comes from 
the $\gamma_5$ matrix in the second term of \geqn{eq:nuMM}.
It is also possible to have interference between the SM and
electromagnetic moment contributions,
\begin{align}
  \overline{\mathcal M_W  \mathcal M^*_{\overset{M}{E}}}
& =
  \frac{8C_{ev}} 3 (2 J_v + 1)
  \sqrt 2 G_F a_{ij} \mu_B
\nonumber 
\\
& \times
\Bigl(\mu_{ij}^*~{\rm or}~\epsilon_{ij}^* \Bigr)
  ( m_j \pm m_i) 
  \left( E_{\nu_i} - \overline E \right),
\label{eq:interference_term}
\end{align}
with $\bar E \equiv (E_{eg}-\omega) [q^2 -  \Delta m^2_{ji}] / 2 q^2$.
However, the interference between magnetic and electric
moment contributions is zero after integrating over $E_{\nu_i}$.

The differential emission rate  
\cite{Dinh:2012qb,Song:2015xaa,Zhang:2016lqp} is,
\begin{eqnarray}\label{eq:diff_xsec}
  \frac {d \Gamma_{ij}} {d E_{\nu_i}}
=
  \frac {\Gamma_0} {(E_{vg} - \omega)^2 \omega}
  \frac {\overline{|\mathcal M_W + \mathcal M_{M,E}|^2}}
        {8 G_F^2 C_{ev} (2J_v + 1)},
\end{eqnarray}
with $E_{vg} \equiv E_v - E_g$.
The reference decay width $\Gamma_0$,
\begin{eqnarray}
  \Gamma_0 
\equiv
  (2 J_v + 1) \frac{n_a^2 C_{ev} G_F^2 |{\bf d}_{gv}\cdot {\bf E}_ 0|^2} \pi,
\end{eqnarray}
regulates the total number of decays.
In practice only a fraction $\eta$ of the total
volume $V$ can be enhanced by $n_a^2 n_\gamma$
where $n_a$ and $n_\gamma$ are the atomic and
photon number densities.

The momentum conservation 
fixes the value of the integration range to be 
$ \bar E - \omega \Delta_{ij} / 2
\leq 
  E_{\nu_i}
\leq 
  \bar E + \omega \Delta_{ij} / 2$
where the relative energy width is
$ \Delta_{ij} 
\equiv
  \sqrt{[q^2  - (m_i + m_j)^2][q^2  - (m_i - m_j)^2]} / q^2$.
Since \geqn{eq:interference_term} is anti-symmetric 
over $E_{\nu_i} - \bar E$, the interference term cannot
survive the neutrino energy integration.
So the total decay rate contains only three parts
$\Gamma = \Gamma_W + \Gamma_{M,E}$ where $\Gamma_W$ is the SM
contribution and $\Gamma_M$ ($\Gamma_E$ ) the magnetic (electric)
moment one. 

\begin{figure}[t]
\centering
\includegraphics[width=0.48\textwidth]{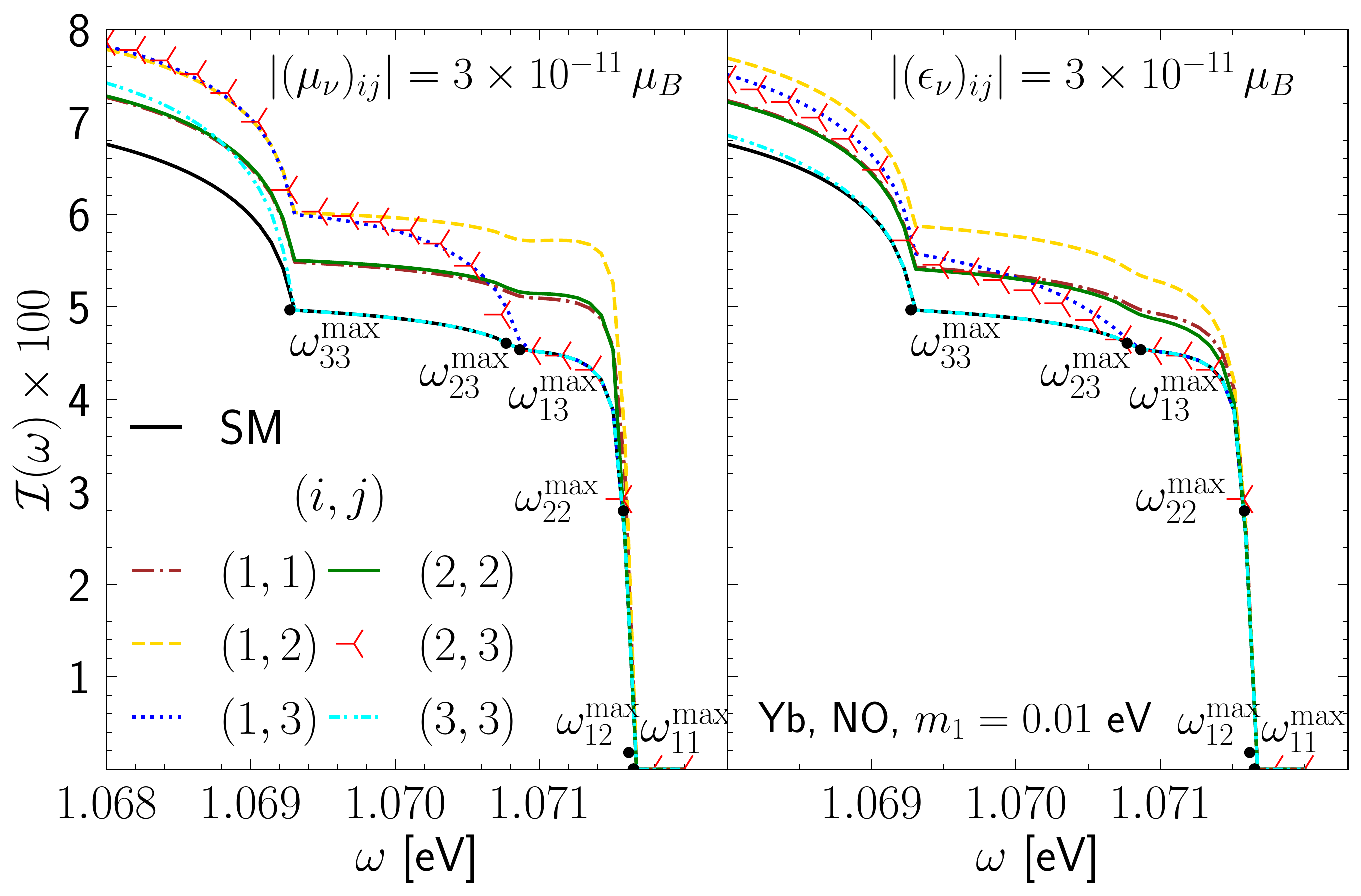}
\caption{
The total spectral function $\mathcal I \equiv 
\mathcal I_W + |(\mu_\nu)_{ij}|^2 \mathcal I_M$ ({\bf left})
or $\mathcal I \equiv 
\mathcal I_W + |(\epsilon_\nu)_{ij}|^2 \mathcal I_E$ ({\bf right})
for the Yb atom as a function 
of the trigger laser frequency $\omega$. The solid black line 
corresponds to vanishing neutrino magnetic/electric moment,
$(\mu_{\nu})_{ij} = (\epsilon_{\nu})_{ij} = 0$, while the colored lines with one non-zero
$(\mu_{\nu})_{ij}$ or $(\epsilon_{\nu})_{ij} = 3\times 10^{-11}\mu_B$ at a time.
The black dots correspond to the kinematic 
thresholds $\omega_{ij}^{\rm max}$. For illustration, we 
take the normal ordering with $m_1 = 0.01$ eV hypothesis.
}
\label{fig:event_distribution}
\end{figure}

Integrating \geqn{eq:diff_xsec} over the neutrino energy
$E_{\nu_i}$ renders the decay rate for the magnetic (electric) moment
to $\Gamma_M \equiv \Gamma_0 |(\mu_\nu)_{ij}| \mathcal I_M$
($\Gamma_E \equiv \Gamma_0 |(\epsilon_\nu)_{ij}| \mathcal I_E$) with,
\begin{align} 
 \mathcal I_{\overset M E}
& \equiv
  \sum_{ij} 
  \left(\frac{\mu_B}{ G_F}\right)^2
     \frac{\omega^2\Delta_{ij}\Theta(\omega - \omega_{ij}^{\rm max})}{9(E_{vg} - \omega)^2}
\label{eq:I_EM}
\\ &\times
\Biggl[
    1
  +
    \frac{(m_i \pm m_j)^2 \pm 4m_i m_j}{q^2}
  -
  2
  \left(\frac{\Delta m_{ji}^2}{q^2}\right)^2
   \Biggl].
\nonumber
\end{align}
For each pair of neutrinos $\nu_i$ and $\nu_j$, the emitted
photon energy has an upper limit $\omega_{ij}^{\rm max}$
\cite{Yoshimura:2011ri,Dinh:2012qb,Fukumi:2012rn},
$\omega_{ij}^{\rm max}
\equiv
 \frac 1 2
 [ E_{eg} - (m_i + m_j)^2 / E_{eg}]$,
as illustrated in \gfig{fig:event_distribution}. Apart from
$E_{eg}$, these frequency thresholds $\omega^{\rm max}_{ij}$
are functions of the neutrino absolute masses $m_i$ and $m_j$.
Smaller mass leads to higher $\omega^{\rm max}_{ij}$.

The final-state photon has the same frequency $\omega$ as the trigger
laser. Tuning the trigger laser frequency allows detailed scan
of the $\mathcal I(\omega)$ function. Especially, scanning the
thresholds at $\omega^{\rm max}_{ij}$ can be used to determine
the neutrino mass hierarchy and the absolute masses 
\cite{Yoshimura:2009wq,Dinh:2012qb,Song:2015xaa,Zhang:2016lqp}. 
This requires tuning the trigger laser with a 
precision of $10^{-5}$\,eV \cite{Dinh:2012qb}.

{\it Searching Strategy and Sensitivity Estimation}
--
Interestingly, the threshold scan can also be used to 
separate the magnetic and electric moments. 
As shown in the left panel of \gfig{fig:event_distribution},
the contribution of $(\mu_\nu)_{ij}$ to the total spectral 
function $\mathcal I \equiv \mathcal I_W + |(\mu_\nu)_{ij}|^2\mathcal I_M$ is non-zero only if $\omega < \omega_{ij}^{\rm max}$. 
In other words, the frequency region $\omega < \omega^{\rm max}_{33}$
receives contribution from all elements $(\mu_\nu)_{ij}$
while the region $\omega^{\rm max}_{33} < \omega < \omega^{\rm max}_{23}$
cannot be affected by $(\mu_\nu)_{33}$. Two independent
measurements below and above $\omega^{\rm max}_{33}$ can
identify a nonzero $(\mu_\nu)_{33}$. Similarly, two
independent measurements in the regions of
$(\omega^{\rm max}_{33}, \omega^{\rm max}_{23})$ and
$(\omega^{\rm max}_{23}, \omega^{\rm max}_{22})$ can
identify $(\mu_\nu)_{23}$. Carrying out this procedure
recursively, all the six $(\mu_\nu)_{ij}$ elements can be identified. 
The process is equivalent for $\epsilon_\nu$. 
For $m_1 = 0.01$\,eV,
we take six trigger laser frequencies $\omega_{i=1 \cdots 6} =$
(1.069, 1.07, 1.0708, 1.0712, 1.0716, 1.07164)\,eV.

In addition,
the sign flip $m_j\rightarrow -m_j$ in \geqn{eq:I_EM}
allows separating the magnetic moment contribution from
the electric one. The difference
$\Delta \mathcal I = \mathcal I_M - \mathcal I_E \propto
12 m_i m_j/q^2$ is relatively significant and can even reach 100\% near
the threshold. With two measurements, one near and another
away from threshold, it is possible to distinguish the magnetic
and electric moments. As a conservative estimation, we assign a
universal extra frequency $\omega_0 = 1.068$\,eV below
$\omega_{33}^{\rm max}$ to resolve the ambiguity.

We try to estimate the sensitivity on the neutrino
electromagnetic moments by taking the conservative setup
with
$n_\gamma = n_a = 10^{21}$\,cm$^{-3}$ \cite{Song:2015xaa,BoyeroGarcia:2015dye}.
The number of photon events for the trigger laser frequency
$\omega_i$ is $N(\omega) \equiv T\, \Gamma_0 \mathcal I(\omega)$,
\begin{eqnarray}\label{eq:numb_events}
    N(\omega)
\approx
  173
  \left(\frac{T}{{\rm day}}\right) 
  \left(\frac{V}{100~{\rm cm}^3}\right)
  \left(\frac{n_a \mbox{ or } n_\gamma}{10^{21} {\rm cm}^{-3}}\right)^3
  \mathcal I(\omega).
\quad
\end{eqnarray}
For an exposure of $T = 10$\,days and a volume of $V = 100$\,cm$^{3}$
that are equally assigned for all the seven frequencies $\omega_i$,
we expect the SM background
events to be $N_{i=0 \cdots 6} \approx (120, 107, 87, 82, 79, 39, 2.5)$,
respectively, with a total of 512 events.

\begin{figure}[t]
\includegraphics[width = 0.5\textwidth]{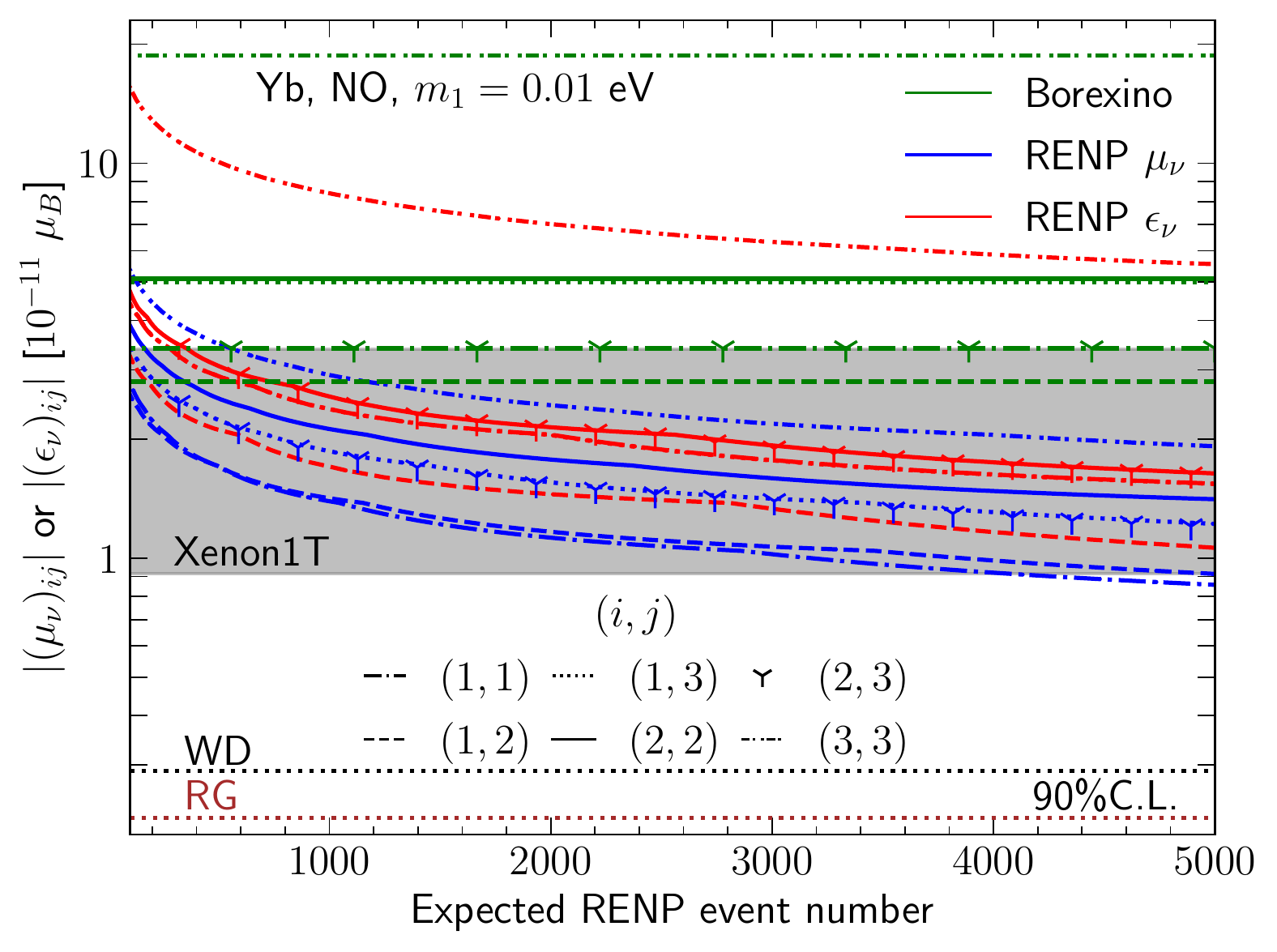}
\caption{
The expected 90\%\,C.L. RENP sensitivity on
the neutrino magnetic moment $(\mu_\nu)_{ij}$ (blue lines)
and electric one $(\epsilon_\nu)_{ij}$ (red lines)
as a function of the expected RENP event number.
For comparison, the green curves show the Borexino sensitivities
\cite{Borexino:2017fbd} while the dotted lines are the 
bounds from stellar cooling of White Dwarf (black)
\cite{Corsico:2014mpa} and Red Giant (red) \cite{Diaz:2019kim}.
The gray region is the 90\% C. L.
neutrino magnetic moment explanation for the Xenon1T 
excess \cite{XENON:2020rca}.
}
\label{fig:sensitivity_plot}
\end{figure}

\gfig{fig:sensitivity_plot} shows the 90\%\,C.L.
sensitivity curves evaluated in Poisson statistics \cite{Ge:2021lur}
versus the expected RENP event number that can be
translated for future experiments with different
configurations as required design targets.
The sensitivity can reach
$(\mu_{\nu})_{ij} < (1.5 \sim 3.5) \times 10^{-11}\,\mu_B$
($(\epsilon_{\nu})_{ij} < (2 \sim 9) \times 10^{-11}\,\mu_B$)
for a conservative number of 500 events
and further touch $(0.8 \sim 2) \times 10^{-12}\,\mu_B$
($(1.1 \sim 5.5) \times 10^{-12}\,\mu_B$)
for 5000 events. 
Among the magnetic moment elements, $(\mu_\nu)_{11}$ 
has the best sensitivity while the worst case is 
$(\mu_\nu)_{33}$ due to two reasons.
First, the $(\mu_\nu)_{33}$ curve is probed at only a single 
frequency, $\omega_1 = 1.0688$\,eV while $(\mu_\nu)_{11}$ contributes 
at all the 6 frequencies. So the event statistic is the
smallest for $(\mu_\nu)_{33}$ and the largest for $(\mu_\nu)_{11}$.
Secondly, the SM background is also larger for lower 
frequency which makes it harder to probe $(\mu_\nu)_{33}$ in
comparison with other parameters. All electric dipole
moments have relatively worse sensitivity than their
magnetic counterparts. This is because the electric 
dipole contribution is suppressed by the sign flip 
of $m_j$. For larger mass, the suppression is larger.
The extreme case happens for the heaviest neutrino with mass
$m_3$ where the sensitivity of $(\epsilon_\nu)_{33}$
is around 2.75 times worse than $(\mu_\nu)_{33}$.

For comparison, we also show the sensitivity of the Borexino 
experiment \cite{Borexino:2017fbd} (green). The result
is translated into neutrino magnetic moments in the mass 
basis by the collaboration using the constraint $\mu^{\rm 
eff}_{\alpha} <2.8\times10^{-11}\mu_B $,
taking $(\epsilon_\nu)_{ij} = 0$ and only one non-zero
$(\mu_\nu)_{ij}$ at a time. Although $(\mu_\nu)_{33}$ still
has the worst sensitivity, the best one occurs for
$(\mu_\nu)_{12} < 2.7\times 10^{-11}\,\mu_B$ instead of
$(\mu_\nu)_{11}$. 
The RENP experiment can exceed this limit for all 
components of $\mu_\nu$ with 1300 events.
The gray band shows the neutrino magnetic moment
explanation to the Xenon1T anomaly,
$\mu_\nu^{\rm eff} \in (0.9 \sim 3.5) \times 
10^{-11}\,\mu_B$ \cite{XENON:2020rca}. Our proposed RENP
setup can probe this region with 500 events. 

For comparison,
the astrophysical constraints have even smaller numbers
at 90\%\,C.L. with $\mu_\nu < 2.9\times 10^{-12}\,\mu_B$
(yellow-dotted) for white dwarfs \cite{Hansen:2015lqa}
and $\mu_\nu < 2.2\times 10^{-12}\,\mu_B$ (red-dotted)
for red giants \cite{Diaz:2019kim}. Although we show
these two sensitivities in \gfig{fig:sensitivity_plot}
for comparison, one needs to keep in mind that there are
various uncertainties for astrophysical measurements.

{\it Conclusions}
--
With $\mathcal O$(eV) momentum transfer, the RENP process
is sensitive to light mediator including the massless photon. 
This feature provides a sensitive probe of the neutrino
electromagnetic moments. The sensitivity can reach
$(1.5 \sim 3.5) \times 10^{-11} \mu_B$ for the magnetic moment
and $(2 \sim 9) \times 10^{-11} \mu_B$ for the electric one
with 500 events.
Further reduction by a factor of 2 is possible with
5000 events. The six components of $\mu_\nu$ or
$\epsilon_\nu$ in the mass basis appear in different frequency
regions which allows frequency scan to
identify each component step-wisely.
Once measured, the different dependence
on the trigger laser frequency allows separation of the magnetic
and electric moments. All these features make the
RENP a unique probe of the neutrino eletromagnetic moments
and the fundamental new physics behind them.

\section*{Acknowledgements}

This work is supported by the Double First Class start-up fund
(WF220442604), the Shanghai Pujiang Program (20PJ1407800),
National Natural Science Foundation of China (No. 12090064),
and Chinese Academy of
Sciences Center for Excellence in Particle Physics (CCEPP).

\end{document}